\documentclass[preprint,aps]{revtex4}
\usepackage{graphicx} 

\newcommand{\pdn}[3]{\frac{\partial^{#3}#1}{\partial#2^{#3}}}
\newcommand{\pd}[2]{\frac{\partial#1}{\partial#2}}

\usepackage{amsmath}

\begin{document}
\title{Coherent instabilities of intense high-energy "white" charged-particle
beams in the presence of nonlocal effects within the context of
the Madelung fluid description}
\author{{\bf R. Fedele}}
\affiliation{\small Dipartimento di Scienze Fisiche,
Universit\`{a} Federico II and INFN Sezione di Napoli, Complesso
Universitario di M.S. Angelo, via Cintia, I-80126 Napoli, Italy}
\author{{\bf D. Anderson}\,\,and\,\,{\bf M. Lisak}}
\affiliation{\small Department of Electromagnetics, Chalmers
University of Technology, G\"oteborg, Sweden}
\begin{abstract}
\small{A hydrodynamical description of coherent instabilities that
take place in the longitudinal dynamics of a charged-particle coasting beam
in a high-energy accelerating machine is presented. This is done in the
framework of the Madelung fluid picture provided by the Thermal Wave Model.
The well known coherent instability charts in the complex plane of the
longitudinal coupling impedance for monochromatic beams are recovered. The
results are also interpreted in terms of the deterministic approach to
modulational instability analysis usually given for monochromatic large
amplitude wave train propagation governed by the nonlinear Schr\"odinger
equation. The instability analysis is then extended to a non-monochromatic
coasting beam with a given thermal equilibrium distribution, thought as a
statistical ensemble of monochromatic incoherent coasting beams ("white"
beam). In this hydrodynamical framework, the phenomenon of Landau damping
is predicted without using any kinetic equation governing the phase space
evolution of the system.}
\end{abstract}

\maketitle

\section{Introduction to the Madelung fluid picture}

A very valuable seminal contribution to quantum mechanics was
given by de Broglie around 1926 with the concept of "quantum
potential", just after proposing his theory of pilot waves
\cite{deBroglie1}. However, an organic presentation of this idea
came only several years later \cite{deBroglie3}. At the beginning
of Fifties, Bohm also have considered the concept of quantum
potential \cite{Bohm}. Actually, the concept was already naturally
appearing in a hydrodynamical description proposed in 1926 by
Madelung \cite{Madelung} (first proposal of a hydrodynamical model
of quantum mechanics), followed by the proposal of Korn in 1927
\cite{Korn}. The Madelung fluid description of quantum mechanics
revealed to be very fruitful in a number of applications: from the
pilot waves theory to the hidden variables theory, from stochastic
mechanics to quantum cosmology (for a historical review, see Ref.
\cite{Auletta}).

In the recently-past years, it has been also applied to
disciplines where the quantum formalism is a useful tool for
describing the evolution of classical systems (quantum-like
systems) or to solve classical nonlinear partial differential
equations \cite{QL-books}.

In the Madelung fluid description, the wave function, say $\Psi$, being a
complex quantity, is represented in terms of modulus and phase which,
substituted in the Schr\"odinger equation, allow to obtain a pair of
nonlinear fluid equations for the "density" $\rho
= |\Psi|^2$ and the "current velocity" ${\bf
V}={\bf\nabla}\mbox{Arg}(\Psi)$: one is the continuity equation
(taking into account the probability conservation) and the other
one is a Navier-Stokes-like motion equation, which contains a
force term proportional to the gradient of the quantum potential,
i.e., $\propto(\nabla^2 |\Psi|)/ |\Psi|=(\nabla^2 \rho^{1/2})/
\rho^{1/2}$. The nonlinear character of these system of fluid
equations naturally allows to extend the Madelung description to systems
whose dynamics is governed by one ore more NLSEs. Remarkably, during the
last four decades, this quantum methodology was imported practically into
all the nonlinear sciences, especially in nonlinear optics
\cite{Akhmanov-et-al}-- \cite{Agrawal} and plasma physics
\cite{Shukla-et-al_PRep}, \cite{Shukla-Stenflo} and it revealed to be very
powerful in solving a number of problems. Let us consider, the following
(1+1)D nonlinear Schr\"odinger-like equation (NLSE):
\begin{equation}
i\alpha{\frac{\partial\Psi}{\partial
s}}~=~-{\frac{\alpha^2}{2}}{\frac{
\partial^2\Psi}{\partial x^2}}~+~U\left[|\Psi|^2\right]\Psi~~~,  \label{nlse}
\end{equation}
where $U\left[|\Psi|^2\right]$ is, in general, a functional of
$|\Psi|^2$, the constant $\alpha$ accounts for the dispersive
effects, and $s$ and $x$ are the timelike and the configurational
coordinates, respectively. Let us assume
\begin{equation}
\Psi~=~\sqrt{\rho (x,s)}\exp\left[{\frac{i}{\alpha}}\Theta
(x,s)\right]~~~, \label{modulus-phase}
\end{equation}
then substitute (\ref{modulus-phase}) in (\ref{nlse}). After
separating the real from the imaginary parts, we get the following
Madelung fluid representation of (\ref{nlse}) in terms of pair of
coupled fluid equations:
\begin{equation}
{\frac{\partial \rho }{\partial s}} + {\frac{\partial }{\partial x
}} \left(\rho V\right) =0\,,  \label{madelung-fluid-continuity}
\end{equation}
(continuity)
\begin{equation}
\left( {\frac{\partial }{\partial s}} + V {\frac{\partial
}{\partial x }} \right)V = - {\frac{\partial U }{\partial x}} +
{\frac{\alpha^2}{2}}{\frac{
\partial}{\partial x}}\left[{\frac{1 }{\rho^{1/2}}} {\frac{\partial^2
\rho^{1/2}}{\partial x^2}}\right],  \label{madelung-fluid-motion}
\end{equation}
(motion) where the current velocity $V$ is given by
\begin{equation}
V(x,s)={\frac{\partial \Theta (x,s)}{\partial x}}\,.
\label{V-definition}
\end{equation}

In order to give the Madelung fluid description of a
charged-particle beam, in the next section, we present the NLSE
describing  the longitudinal dynamics of a coasting beam in the
presence of nonlinear collective and nonlocal effects in
high-energy accelerating machines in the framework of the Thermal
Wave Model (TWM).

\section{The NLSE in the framework of TWM.}

Within the TWM framework, the longitudinal dynamics of particle bunches is
described in terms of a complex wave function $\Psi(x, s)$, where $s$ is
the distance of propagation and $x$ is the longitudinal extension of the
particle beam, measured in the moving frame of reference.  The particle
density, $\lambda(x, s)$, is related to the wave function according to
$\lambda(x, s) = |\Psi(x, s)|^2$, \cite{Fedele-Miele}. The collective
longitudinal evolution of the beam in a circular high-energy accelerating
machine is governed by the Schr\"odinger-like equation
\begin{equation}
i \epsilon \pd{\Psi}{s} + \frac{\epsilon^2 \eta}{2}
\pdn{\Psi}{x}{2} + U(x, s) \Psi = 0\,\,, \label{TWM-eq}
\end{equation}
where $\epsilon$ is the longitudinal beam emittance and $\eta$ is
the slip factor, \cite{Lawson_1988}, defined as $\eta =
\gamma_{T}^{-2} - \gamma^{-2}$ ($\gamma_T$ being the transition
energy, defined as the inverse of the momentum compaction,
\cite{Lawson_1988}, and $\gamma$ being the relativistic factor);
$U(x, s)$ is the effective dimensionless (with respect to the
nominal particle energy, $E_0 = m \gamma c^2$) potential energy
given by the interaction between the bunch and the surroundings.
Note that $\eta$ can be positive (above transition energy) or
negative (below transition energy).  Above the transition energy,
in analogy with quantum mechanics, $1/\eta$ plays the role of an
effective mass associated with the beam as a whole. Below
transition energy, $1/\eta$ plays the role of a ``negative mass''.

Equation~(\ref{TWM-eq}) has to be coupled with an equation for
$U$. If no external sources of electromagnetic fields are present
and the effects of charged-particle radiation damping is
negligible, the self-interaction between the beam and the
surroundings, due to the image charges and the image currents
originated on the walls of the vacuum chamber, makes $U$ a
functional of the beam density. It can proven that, in a
torus-shaped accelerating machine, characterized by a toroidal
radius $R_0$ and a poloidal radius $a$, for a coasting beam of
radius $b<<a$ travelling at velocity $\beta c$ ($\beta \leq 1$ and
$c$ being the speed of light), the self-interaction potential
energy is given by \cite{Johannisson-et-al_PRE} (a more general
expression is given in Ref. \cite{Schamel_2000_pp}):
\begin{equation}
U[\lambda_1 (x, s)] = \quad \frac{q^2 \beta c}{E_0} \left( R_0
Z_{I}^{\prime} \lambda_1 (x, s) + Z_{R}^{\prime} \int_{0}^{x}
\lambda_1 (x^{\prime}, s) \, dx^{\prime} \right),
\label{functional-U}
\end{equation}
where $\lambda_1 (x,s)$ is an (arbitrarily large) line beam
density perturbation, $q$ is the charge of the particles,
$\epsilon_0$ is the vacuum dielectric constant, $Z_{R}^{\prime}$
and $Z_{I}^{\prime}$ are the resistive and the total reactive
parts, respectively, of the longitudinal coupling impedance per
unit length of the machine. Thus, the coupling impedance per unit
length can be defined as the complex quantity $Z^{'} = Z_R^{'} + i
Z_I^{'}$. In our simple model of a circular machine, it is easy to
see that \cite{Lawson_1988}, \cite{Schamel_2000_pp}:
\begin{equation}
Z_{I}^{\prime} = \frac{1}{2 \pi R_0} \left( \frac{g_0 Z_0}{2 \beta
\gamma^2} - \omega_0 \mathcal{L} \right) \equiv \frac{Z_I}{2 \pi
R_0}, \label{reactance}
\end{equation}
where $Z_0$ is the vacuum impedance, $\omega_0 = \beta c/R_0$ is
the nominal orbital angular frequency of the particles and
$\mathcal{L}$ is the total inductance.  This way, $Z_I$ represents
the total reactance as the difference between the total space
charge capacitive reactance, $g_0 Z_0/(2 \beta \gamma^2)$, and the
total inductive reactance, $\omega_0 \mathcal{L}$. Consequently,
in the limit of negligible resistance, Eq.~(\ref{functional-U})
reduces to
\begin{equation}
U[\lambda_1] = \frac{q^2 \beta c}{2 \pi E_0} \left( \frac{g_0
Z_0}{2 \beta \gamma^2} - \omega_0 \mathcal{L} \right) \lambda_1.
\label{functional-U-1}
\end{equation}
By definition, an unperturbed coasting beam has the particles
uniformly distributed along the longitudinal coordinate $x$.
Denoting by $\rho(x,s)$ the line density and by $\rho(x, 0)$ the
unperturbed one, in the TWM framework we have the following
identifications: $\rho (x,s) = |\Psi (x,s)|^2$, $\rho_0 = |\Psi(x,
0)|^2 \equiv |\Psi_0|^2$, where $\Psi_0$ is a complex function
and, consequently, $\lambda_1 (x, s)= |\Psi(x, s)|^2 -
|\Psi_0|^2$. Thus, the combination of Eq.~(\ref{TWM-eq}) and
Eq.~(\ref{functional-U}) gives the following evolution equation
for the beam
\begin{equation}
i{\partial \Psi\over\partial s}~+~{\alpha \over 2}{\partial^2
\Psi\over\partial x^2}~+~{\cal
X}\left[|\Psi|^2~-~|\Psi_0|^2\right]\Psi~+~{\cal R}~\Psi~\int_0^x~
\left[|\Psi (x',s)|^2~-~|\Psi_0|^2\right]~dx'~=~0\,,
\label{schrodinger-like-bis}
\end{equation}
where
\begin{eqnarray}
\alpha &=&
\epsilon\eta=\epsilon\left(\gamma^{-2}-\gamma_{T}^{-2}\right),
\label{alpha}\\
{\cal X} &=& {q^2 \beta c R_0\over \epsilon E_0} Z_I^{\prime},\label{cal-X}\\
{\cal R} &=& {q^2 \beta c\over \epsilon E_0}
Z_R^{\prime}.\label{cal-R}
\end{eqnarray}
Equation~(\ref{schrodinger-like-bis}) belongs to the family of
NLSEs governing the propagation and dynamics of wave packets in
the presence of nonlocal effects. The modulational instability of
such an integro-differential equation has been investigated for
the first time in literature in Ref. \cite{Anderson-et-al_PLA}.
Some nonlocal effects associated with the collective particle beam
dynamics have been recently described with this equation. Note
that Eq. (\ref{schrodinger-like-bis}) can be cast in the form of
Eq. (\ref{nlse}), provided that (\ref{alpha})-(\ref{cal-R}) are
taken and the following expression for the nonlinear potential is
assumed, i.e.,
\begin{equation}
U[|\Psi|^2]= -\alpha \left\{{\cal
X}\left[|\Psi|^2~-~|\Psi_0|^2\right]~+~{\cal R}~\int_0^x~
\left[|\Psi (x',s)|^2~-~|\Psi_0|^2\right]~dx'\right\}\,\,.
\label{effective-potential}
\end{equation}

\section{Coherent instability analysis and its identification with
the modulational instability}

\subsection{Deterministic approach to MI (monochromatic coasting beam)}
\begin{figure}
\includegraphics[width=0.42\linewidth]{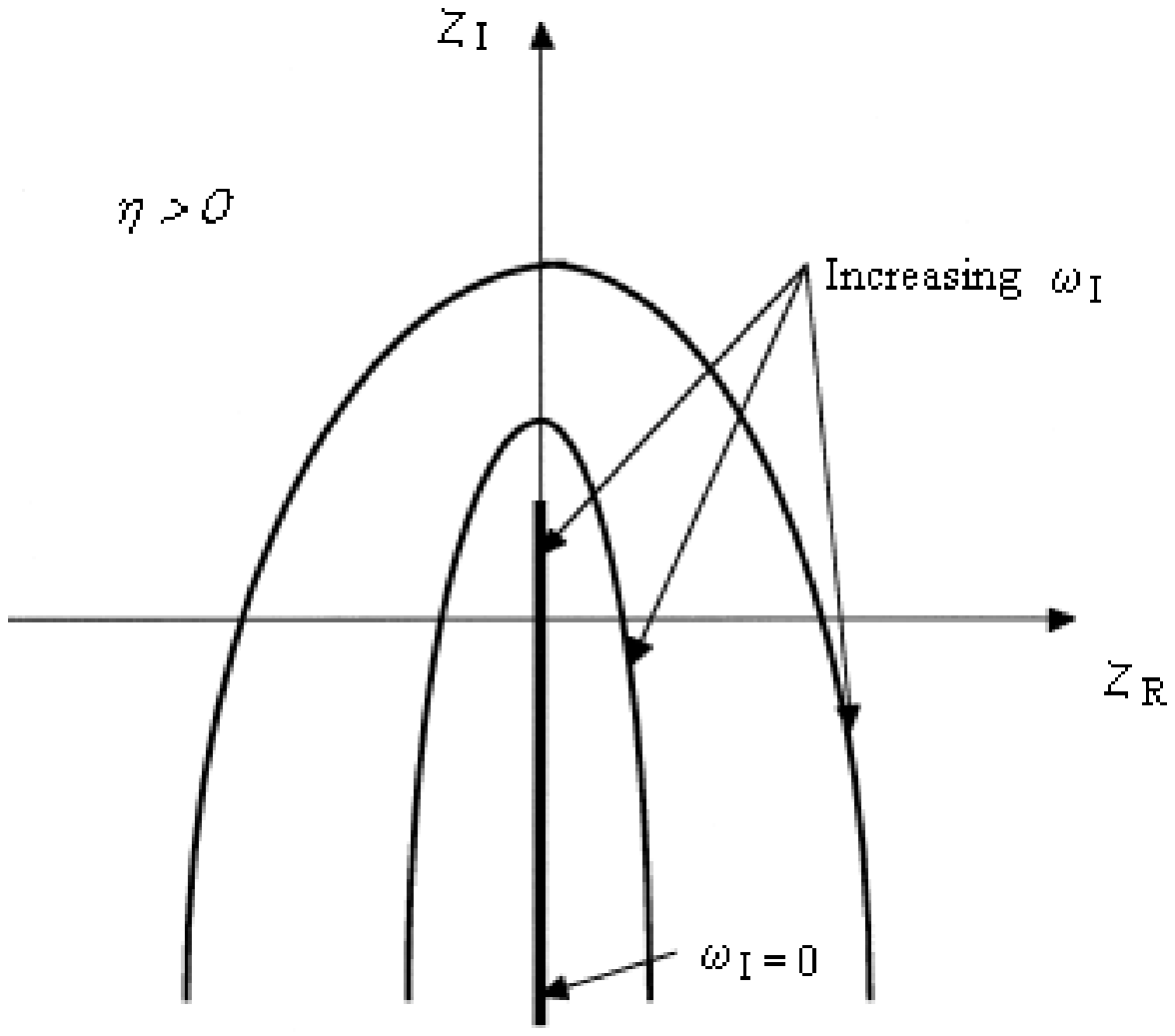}
\includegraphics[width=0.42\linewidth]{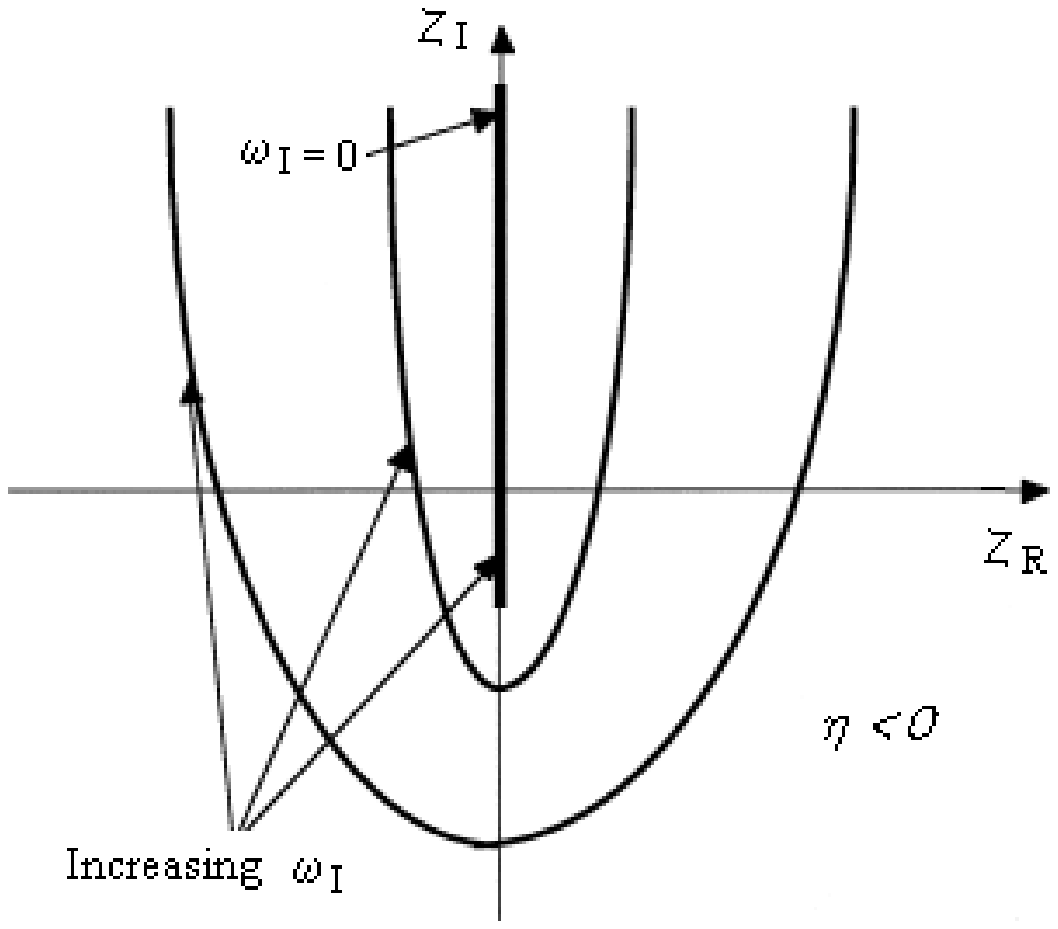}
\caption{\label{figure2}{\it Qualitative plots of the modulational
instability curves in the plane $({\cal Z}_R - {\cal Z}_I)$ of a
coasting beam below the transition energy, ($\eta
> 0$) and above the transition energy,
($\eta <0 $), respectively. The bold face vertical straight lines
represent the stability region ($\omega_I =0$)}.}
\end{figure}

Under the conditions assumed above, let us consider a
monochromatic coasting beam travelling in a circular high-energy
machine with the unperturbed velocity $V_0$ and the unperturbed
density $\rho_0 =|\Psi_0|^2$ (equilibrium state). In these
conditions, all the particles of the beam have the same velocity
and their collective interaction with the surroundings is absent.
In the Madelung fluid representation, the beam can be thought as a
fluid with both current velocity and density (i.e., $\rho_0$)
uniform and constant. In this state, the Madelung fluid equations
(\ref{madelung-fluid-continuity}) and
(\ref{madelung-fluid-motion}) vanish identically. Let us now
introduce small perturbations in $V(x,s)$ and $\rho (x,s)$, i.e.,
\begin{eqnarray}
V &=& V_0 + V_1\,, \,\,\,\,\,\,|V_1|\,<<\,|V_0|\,,\label{V}\\
\rho &=& \rho_0 + \rho_1\,,
\,\,\,\,\,\,|\rho_1|\,<<\,\rho_0\,.\label{rho}
\end{eqnarray}
By introducing (\ref{V}) and (\ref{rho}) in the pair of equations
(\ref{madelung-fluid-continuity}) and
(\ref{madelung-fluid-motion}), after linearizing, we get the
following system of equations:
\begin{eqnarray}
&&{\partial\rho_1\over\partial s} + V_0
{\partial\rho_1\over\partial x} + \rho_0 {\partial
V_1\over\partial x} = 0
\,,\label{perturbed-continuity}\\
&&{\partial V_1\over\partial s} + V_0 {\partial V_1\over\partial
x} = \alpha{\cal R}\rho_1 + \alpha{\cal
X}{\partial\rho_1\over\partial x} + {\alpha^2\over
4\rho_0}{\partial^3\rho_1\over\partial x^3}
\,.\label{perturbed-motion}
\end{eqnarray}
In order to find the linear dispersion relation, we take the
Fourier transform of the system of equations
(\ref{perturbed-continuity}) and (\ref{perturbed-motion}), i.e. we
express the quantities $\rho_1(x,s)$ and $V_1 (x,s)$ in terms of
their Fourier transforms $\tilde{\rho_1}(k,\omega)$ and
$\tilde{V_1}(k,\omega)$, respectively,
\begin{eqnarray}
\rho_1 (x,s) =
\int\,dk\,d\omega\,\tilde{\rho_1}(k,\omega)e^{ikx-i\omega s}\,,
\label{Fourier-rho-1}\\
V_1 (x,s) = \int\,dk\,d\omega\,\tilde{V_1}(k,\omega)e^{ikx-i\omega
s}\,,\label{Fourier-V-1}
\end{eqnarray}
and, after substituting in (\ref{perturbed-continuity}) and
(\ref{perturbed-motion}), we get the following system of algebraic
equations:
\begin{eqnarray}
-\rho_0 k \tilde{V_1} &=& \left(k V_0 -
\omega\right)\tilde{\rho_1}\,,
\label{Fourier-continuity}\\
i\left(k V_0 - \omega\right)\tilde{V_1} &=& \left(\alpha {\cal R}+
i\alpha k{\cal X} -
i{\alpha^2\over4\rho_0}k^3\right)\tilde{\rho_1}\,.\label{Fourier-motion}
\end{eqnarray}
By combining (\ref{Fourier-continuity}) and (\ref{Fourier-motion})
we finally get the dispersion relation
\begin{equation}
\left({\omega\over k} - V_0\right)^2 = i \alpha\rho_0 \left({{\cal
Z}\over k}\right) + {\alpha^2 k^2\over 4} \,,
\label{dispersion-relation-1}
\end{equation}
where we have introduced the complex quantity ${\cal Z}={\cal R} +
i k{\cal X}\equiv {\cal Z}_R + i {\cal Z}_I$, proportional to the
longitudinal coupling impedance per unity length of the beam. In
general, in Eq. (\ref{dispersion-relation-1}), $\omega$ is a
complex quantity, i.e., $\omega \equiv \omega_R + i \omega_I$. If
$\omega_I \neq 0$, the modulational instability takes place in the
system. Thus, by substituting the complex form of $\omega$ in Eq.
(\ref{dispersion-relation-1}), separating the real from the
imaginary parts and using (\ref{alpha}), we finally get:
\begin{equation}
{\cal Z}_I =- \eta{\epsilon k \rho_0 \over 4\omega_I^2}{\cal
Z}_R^2 + {1\over \eta}{\omega_I^2\over\epsilon k \rho_0 } +
\eta{\epsilon k^3\over 4\rho_0} \,. \label{impedances-eq}
\end{equation}
This equation fixes, for any values of the wavenumber $k$ and any
values of the growth rate $\omega_I$ a relationship between real
and imaginary parts of the longitudinal coupling impedance. For
each $\omega_I \neq 0$, running the values of the slip factor
$\eta$, it describes two families of parabolas in the complex
plane $({\cal Z}_R\,-\,{\cal Z}_I)$. Each pair $({\cal Z}_R,{\cal
Z}_I)$ in this plane represents a working point of the
accelerating machine. Consequently, each parabola is the locus of
the working points associated with a fixed growth rate of the MI.
According to Figure \ref{figure2}, below the transition energy
($\gamma < \gamma_T$),\,$\eta$ is positive and therefore the
instability parabolas have a negative concavity, whilst above the
transition energy ($\gamma > \gamma_T$), since $\eta$ is negative
the instability parabolas have a positive concavity (negative mass
instability).
It is clear from Eq. (\ref{impedances-eq}) that, approaching
$\omega_I =0$ from positive (negative) values, the two families of
parabolas reduce asymptotically to a straight line upper (lower)
unlimited located on the imaginary axis. The straight line
represent the only possible region below (above) the transition)
energy where the system is modulationally stable against small
perturbations in both density and velocity of the beam, with
respect to their unperturbed values $\rho_0$ and $V_0$,
respectively (note that density and velocity are directly
connected with amplitude and phase, respectively, of the wave
function $\Psi$). Any other point of the complex plane belongs to
a instability parabola ($\omega_I\neq 0$).

In the limit of small dispersion, i.e., $\epsilon k <<1$, the
second term of the right hand side of Eq.
(\ref{dispersion-relation-1}) can be neglected and Eq.
(\ref{impedances-eq}) reduces to
\begin{equation}
{\cal Z}_I \approx - \eta{\epsilon k \rho_0 \over
4\omega_I^2}{\cal Z}_R^2 + {1\over \eta}{\omega_I^2\over\epsilon k
\rho_0 } \,. \label{impedances-eq-1}
\end{equation}
Furthermore, for purely reactive impedances (${\cal Z}_R\equiv
0$), Eq. (\ref{schrodinger-like-bis}) reduces to the cubic NLSE
and the corresponding dispersion relation gives (note that in this
case $\omega_R = V_0 k$)
\begin{equation}
{\omega_I^2\over k^2} = -\epsilon\eta\rho_0 \left({{\cal Z}_I\over
k}\right) + {\alpha^2 k^2\over 4} \,, \label{omega_I-relation}
\end{equation}
from which it is easily seen that the system is modulationally
unstable ($\omega_I^2 >0$) under the following conditions
\begin{eqnarray}
\eta\,{\cal Z}_I&\,>\,&0 \label{MI-condition-1}\\
\rho_o &\,>\,& {\epsilon\eta k^2\over 4{\cal X}_I}\,.\label{MI-2}
\end{eqnarray}
Condition (\ref{MI-condition-1}) is a well known coherent
instability condition for purely reactive impedances which
coincides with the well known "Lighthill criterion"
\cite{Lighthill} associated with the cubic NLSE. This aspects has
been pointed out for the first time in Ref.s
\cite{Fedele-et-al_EPAC92}, \cite{Fedele-et-al_PLA}.
\bigskip
\begin{table}[ht]
\begin{tabular}{c|c|c}
{} & {${\cal Z}_I\,>\,0$} & \,\,\,\,{${\cal Z}_I\,<\,0$}\\
{} & {\small (capacitive)} & \,\,\,\,\,{\small (inductive)}\\
\hline
& &\\
$\eta\,>\,0$\,\,\,\,\,\, & {\bf stable} &
\,\,\,\,\,{\bf unstable}\\
{\small(below transition energy)} & {} & \,\,\,\,\,{}\\
\hline
& & \\
$\eta\,<\,0$\,\,\,\,\,\, & \,\,\,\,{\bf unstable}\,\,\,\, &
\,\,\,\,\,{\bf stable}\\
{\small(above transition energy)} & {} & \,\,\,\,\,{}\\
\end{tabular}
\medskip
\caption{\label{Table2}{\it Coherent instability scheme of a
monochromatic coasting beam in the case of a purely reactive
impedance} $\left({\cal Z}_R =0\right)$. {\it The instability
corresponding to $\eta <0$ is usually referred to as "negative
mass instability".}}
\end{table}
According to Table I, this condition implies that the system is
modulationally unstable below (above) transition energy and for
capacitive (inductive) impedances and stable in the other
different possible circumstances.

Condition (\ref{MI-2}) implies that the instability threshold is
given by the nonzero minimum intensity $\rho_{0m}=\epsilon\eta
k^2/4{\cal X}_I$.

\subsection{MI analysis of a white coasting beam} The
dispersion relation (\ref{dispersion-relation-1}) allows to write
an expression for the admittance of the coasting beam ${\cal
Y}\equiv 1/{\cal Z}$:
\begin{equation}
k{\cal Y} = {i\alpha\rho_0\over \left(\omega/k - V_0\right)^2 -
\alpha^2 k^2/ 4} \,. \label{dispersion-relation-2}
\end{equation}
Let us now consider a non-monochromatic coasting beam. Such a
system may be thought as an ensemble of incoherent coasting beams
with different unperturbed velocities (white beam). Let us call
$f_0 (V)$ the distribution function of the velocity at the
equilibrium. The subsystem corresponding to a coasting beam
collecting the particles having velocities between $V$ and $V+dV$
has an elementary admittance $d{\cal Y}$. Provided, in Eq.
(\ref{dispersion-relation-2}), to replace $\rho_0$ with
$f_0(V)dV$, the expression for the elementary admittance is easily
given:
\begin{equation}
kd{\cal Y} = {i\alpha \,f_0(V)\,dV\over \left(V - \omega/k
\right)^2 - \alpha^2 k^2/ 4} \,.
\label{elementary-dispersion-relation-2}
\end{equation}
All the elementary coasting beams in which we have divided the
system suffer the same electric voltage per unity length along the
longitudinal direction. This means that the total admittance of
the system is the sum of the all elementary admittances, as it
happens for a system of electric wires connected all in parallel.
Therefore,
\begin{equation}
k{\cal Y} = i\alpha \,\int\,{f_0(V)\,dV\over \left(V -\omega/k
\right)^2 - \alpha^2 k^2/ 4} \,. \label{dispersion-relation-3}
\end{equation}
Of course, this dispersion relation can be cast also in the
following way:
\begin{equation}
1 = i\alpha \left({{\cal Z}\over k}\right)
\,\int\,{f_0(V)\,dV\over \left(V -\omega/k \right)^2 - \alpha^2
k^2/ 4} \,, \label{dispersion-relation-4}
\end{equation}
where we have introduced the total impedance of the system which
is the inverse of the total admittance, i.e., ${\cal Z}=1/{\cal
Y}$.

An interesting equivalent form of Eq.
(\ref{dispersion-relation-4}) can be obtained. To this end, we
first observe that the folowing identity holds:
$$
{1\over \left(V - \omega/k \right)^2 - \alpha^2 k^2/ 4} =
{1\over\alpha k}\,\left[{1\over \left(V-\alpha k/2\right)-\omega
/k} - {1\over \left(V + \alpha k/2\right)-\omega /k}\right]\,.
$$
Then, using this identity in Eq. (\ref{dispersion-relation-4}) it
can be easily shown that:
\begin{equation}
1 = i\left({{\cal Z}\over k}\right){1\over k}
\,\left[\int\,{f_0(V)\,dV\over \left(V-\alpha k/2\right)-\omega/k}
- \int\,{f_0(V)\,dV\over \left(V+\alpha
k/2\right)-\omega/k}\right]\,, \label{dispersion-relation-5}
\end{equation}
which, after defining the variables $p_1 = V-\alpha k/2$ and $p_2
= V+\alpha k/2$, can be cast in the form:
\begin{equation}
1 = i\left({{\cal Z}\over k}\right){1\over k}
\,\left[\int\,{f_0(p_1 + \alpha k/2)\,dp_1\over p_1-\omega/k} -
\int\,{f_0(p_2 - \alpha k/2)\,dp_2\over p_2-\omega/k}\right]\,,
\label{dispersion-relation-6}
\end{equation}
and finally in the following form:
\begin{equation}
1 = i\alpha\left({{\cal Z}\over k}\right)\,\int\,{f_0(p + \alpha
k/2) - f_0(p - \alpha k/2)\over \alpha k}\, {dp\over p-\omega
/k}\,. \label{dispersion-relation-7}
\end{equation}
We soon observe that, assuming that $f_0 (V)$ is proportional to
$\delta (V-V_0)$, from Eq. (\ref{dispersion-relation-7}) we easily
recover the dispersion relation for the case of a monochromatic
coasting beam (see Eq. (\ref{dispersion-relation-1})\,). In
general, Eq. (\ref{dispersion-relation-7}) takes into account the
equilibrium velocity (or energy) spread of the beam particles, but
it has not obtained with a kinetic treatment. We have only assumed
the existence of an equilibrium state associated with an
equilibrium velocity distribution, without taking into account any
phase-space evolution in terms of a kinetic distribution function.
Our result has been basically obtained within the framework of
Madelung fluid description, extending the standard MI analysis for
monochromatic wave trains to non-monochromatic wave packets
(statistical ensemble of monochromatic coasting beams).

Nevertheless, Eq. (\ref{dispersion-relation-7}) can be also
obtained within the kinetic description provided by the
Moyal-Ville-Wigner description \cite{Wigner} - \cite{Ville}, as it
has been done for the first time in the context of the TWM
\cite{Anderson-et-al_INFN_Prep} soon extended to nonlinear optics
\cite{Fedele-Anderson}-\cite{Helczynski-et-al_IEEE}, plasma
physics \cite{Fedele-et-al_PLA1}, \cite{Marklund}, surface gravity
waves \cite{Onorato-et-al_PRE}, in lattice vibrations physics
(molecular crystals) \cite{Visinescu-Grecu},
\cite{Grecu-Visinescu}.

From the above investigations, and according to the former quantum
kinetic approaches to nonlinear systems \cite{Klimontovich-Silin},
\cite{Alber}, we can summarize the following general conclusions.

\begin{itemize}
\item There are two distinct ways to describe MI. The first, and
the most used one, is a "deterministic" approach, whilst the
second one is a "statistical approach".

\item In the statistical approach, the basic idea is to transit
from the configuration space description, where the NLSE governs
the particular wave-envelope propagation, to the phase space,
where an appropriate kinetic equation is able to show a random
version of the MI. This has been accomplished by using the
mathematical tool provided by the "quasidistribution" (Fourier
transform of the density matrix) that is widely used for quantum
systems. In fact, for any nonlinear system, whose dynamics is
governed by the NLSE, one can introduce a two-points correlation
function which plays the role similar to the one played by the
density matrix of a quantum system \cite{Landau1}-\cite{Weyl}.
Consequently, the governing kinetic equation is nothing but a sort
of nonlinear von Neumann-Weyl equation. In the statistical
approach to modulational instability, a linear stability analysis
of the von Neumann-Weyl equation leads to a phenomenon fully
similar to the well known Landau damping, predicted by L.D. Landau
in 1946 for plasma waves \cite{Landau}

\item The deterministic MI can be recovered for the case of a
monochromatic wavetrain; in particular, it coincides with coherent
instability of a coasting beam in the limit of weak dispersion.

\item A Landau--type damping for a non-monochromatic wavepacket is
predicted and the weak Landau damping is recovered for weak
dispersion, in particular for plasma waves and particle beams in
the usual kinetic Vlasov-Maxwell framework.

\item The interplay between Landau damping and MI characterizes
the statistical behavior of the nonlinear collective wave packet
propagation governed by the NLSE.

\end{itemize}

All the above conclusions have been obtained within the kinetic
description from a dispersion relation fully similar to Eq.
(\ref{dispersion-relation-7}). Consequently, it is absolutely
evident that all the above conclusions can be obtained within the
framework of the Madelung description of a white intense
charged-particle coasting beam. This proves that the Madelung
fluid description of the MI of an ensemble of incoherent beams
(white beam) is equivalent to the one provided by the
Moyal-Ville-Wigner kinetic theory.

\section{Conclusions and Remarks}
In this paper, we have developed a hydrodynamical description of
coherent instability of an intense white coasting charged-particle
beam in high-energy accelerator in the presence of nonlinear
collective and nonlocal effects. The analysis has been based on
the Madelung fluid model within the framework of the TWM. It has
been shown that this quantum hydrodynamical description of MI,
with both deterministic or statistical character, is fully
equivalent to the one provided by the quantum kinetic theory.
Remarkably, the proposed hydrodynamical description is certainly
very convenient in particle accelerators because it is very close
to the standard classical picture of particle beams (in particular
white beams) in particle accelerators.


\begin{thebibliography}{999}

\bibitem{deBroglie1}L. de Broglie, {\it Comptes Rendus \'{a} l'Academie des Sciences}
{\bf 184}, 273 (1927); {\bf 185}, 380 (1927); {\it Journal de
Physique} {\bf 8}, 255 (1927).

\bibitem{deBroglie3}L. de Broglie, {\it Une tentative d'Interpretation Causale et
Non-line\'{a}re de la Meccanique Ondulatoire} (Gauthier-Villars,
Paris, 1956).

\bibitem{Bohm}D. Bohm, {\it Phys. Rev.} {\bf 85}, 166 (1952).

\bibitem{Madelung} E. Madelung, {\it Z. Phys.} {\bf 40}, (1926) 332.

\bibitem{Korn}A. Korn, {\it Zeitschrift f\"{u}r Physik} {\bf 44}, 745 (1927).

\bibitem{Auletta}G. Auletta, {\it Foundation and Interpretation of
Quantum Mechanics} (World Scientific, Singapore, 2000).

\bibitem{QL-books}See for instance, R. Fedele and P.K. Shukla (editors),
{\it Quantum-like Models and Coherent Effects} (World Scientific,
Singapore, 1995), Proc. of the 27th Workshop of the INFN
Eloisatron Project, Erice, Italy 13-20 June 1994; S. Marticno, S.
De Nicola, S. De Siena, R. Fedele and G. Miele (editors), {\it New
Perspectives in the Physics of Mesoscopic Systems} (World
Scientifc, Singapore, 1997), Proc. of the Workshop "New
Perspectives in the Physics of Mesoscopic Systems: Quantum-like
Description and Macroscopic Coherence Phenomena", Caserta, Italy,
18-20 April 1996; P. Chen (editor), {\it Quantum Aspects of Beam
Physics} (World Scientific, Singapore, 1999), Proc. of the
Advanced ICFA Beam Dynamics Workshop on "Quantum Aspects of Beam
Physics", Monterey, California (USA), 4-9 January 1998; P. Chen
(editor), {\it Quantum Aspects of Beam Physics} (World Scientific,
Singapore, 2002), Proc. of the 18th Advanced ICFA Beam Dynamics
Workshop on "Quantum Aspects of Beam Physics", Capri, Italy, 15-20
October 2000; P. Chen and K. Reil (editors), {\it Quantum Aspects
of Beam Physics} (World Scientific, Singapore, 2004), Proc. of the
Joint 28th ICFA Advanced Beam Dynamics and Advanced and Novel
Accelerator Workshop on "Quantum Aspects of Beam Physics",
Hiroshima, Japan, 7-11 January 2003; P.K. Shukla and L. Stenflo
(editors), {\it New Frontiers in Nonlinear Sciences}, Proc. of the
Inter. Topical Conf. on Plasma Physics, Univ. do Algarve, Faro,
Portugal, 6-10 September, 1999, published in {\it Physica Scripta}
{\bf T84} (2000); P.K. Shukla and L. Stenflo (editors), {\it New
Plasma Horizons}, Proc. of the Inter. Topical Conf. on Plasma
Physics, Univ. do Algarve, Faro, Portugal, 3-7 September, 2001,
published in {\it Physica Scripta} {\bf T98} (2002).

\bibitem{Akhmanov-et-al}S.A. Akhmanov, A.P. Sukhuorukov, and R.V. Khokhlov,
{\it Sov. Phys. Usp.} {\bf 93}, 609 (1968).

\bibitem{Shen}Y.R. Shen, {\it The Principles of Nonlinear Optics}
(Wiley-Interscience Publication, New York, 1984).

\bibitem{Agrawal}G.P. Agrawal, {\it Nonlinear Fibre Optics} (Academic Press, San Diego
1995).

\bibitem{Shukla-et-al_PRep}P.K. Shukla, N.N. Rao, M.Y. Yu and N.L. Tasintsadze,
{\it Phys. Rep.} {\bf 138}, 1 (1986).

\bibitem{Shukla-Stenflo}P.K. Shukla and L. Stenflo (editors),
{\it Modern Plasma Science}, Proc. of the Int. Workshop on
Theoretical Plasma Physics, Abdus Salam ICTP, Trieste, Italy, July
5-16, 2004, in {\it Physica Scripta} {T116} (2005).

\bibitem{Fedele-Miele}R.Fedele and G. Miele, { \it Il Nuovo Cimento} {\bf D13}, 1527 (1991).

\bibitem{Lawson_1988}J. Lawson, {\it The Physics of Charged Particle Bea
ms} (Clarendon, Oxford, 1988), 2nd ed.

\bibitem{Johannisson-et-al_PRE}P. Johannisson, D. Anderson, M. Lisak, M. Marklund, R. Fedele
and A. Kim, {\it Phys. Rev. E.} {\bf 69}, 066501 (2004).

\bibitem{Schamel_2000_pp}H. Schamel and R. Fedele, {\it Phys. Plasmas}
{\bf 7}, 3421 (2000).

\bibitem{Anderson-et-al_PLA}D. Anderson, R Fedele, V. Vaccaro, M. Lisak, A. Berntson,
S. Johanson, {\it Phys. Lett. A} {\bf 258}, 244 (1999).

\bibitem{Lighthill}M.J. Lighthill, {\it J. Inst. Math. Appl.} {\bf 1}, 269 (1965); Proc. Roy.
Soc. {\bf 229}, 28 (1967).

\bibitem{Fedele-et-al_EPAC92}R. Fedele, L. Palumbo and V.G. Vaccaro, {\it A Novel
Approach to the Nonlinear Longitudinal Dyanamics in Particle
Accelerators}, Proc. of the Third European Particle Accelerator
Conference (EPAC 92), Berlin, 24-28 March, 1992 edited by H.
Henke, H. Homeyer and Ch. Petit-Jean-Genaz (Edition Frontieres,
Singapore, 1992), p. 762.

\bibitem{Fedele-et-al_PLA}R. Fedele, G. Miele, L. Palumbo and V.G. Vaccaro, {\it Phys. Lett. A}
{\bf 179}, 407 (1993).

\bibitem{Anderson-et-al_INFN_Prep}D. Anderson, R. Fedele, V.G. Vaccaro, M. Lisak,
A. Berntson, S. Johansson, {\it Quantum-like Description of
Modulational and Instability and Landau Damping in the
Longitudinal Dynamics of High-Energy Charged-Particle Beams},
Proc. of  1998 ICFA Workshop on "Nonlinear Collective Phenomena in
Beam Physics". Arcidosso, Italy, September 1-5, 1998, S.
Chattopadhyay, M. Cornacchia, and C. Pellegrini (Ed.s), (AIP
Press, New York, 1999) p.197; D. Anderson, R. Fedele, V.G.
Vaccaro, M. Lisak, A. Berntson, S. Johansson, {\it Modulational
Instabilities and Landau damping within the Thermal Wave Model
Description of High-Energy Charged-Particle Beam Dynamics},
INFN/TC-98/34, 24 November (1998); R. Fedele, D. Anderson, and M.
Lisak, {\it Role of Landau damping in the Quantum-Like Theory of
Charged-Particle Beam Dynamics}, Proc. of Seventh European
Particle Accelerator Conference (EPAC2000), Vienna, Austria, 26-30
June, 2000, p.1489.

\bibitem{Fedele-Anderson}R. Fedele and D. Anderson, {\it J. Opt. B: Quantum Semiclass.
Opt.}, {\bf 2}, 207 (2000).

\bibitem{Fedele-et-al_PS1}R. Fedele, D. Anderson and M. Lisak,
{\it Physica Scripta} {\bf T84}, 27 (2000).

\bibitem{Hall-et-al_PRE}B. Hall, M. Lisak, D. Anderson, R. Fedele,
and V.E. Semenov, {\it Phys. Rev. E}, {\bf 65}, 035602(R) (2002).

\bibitem{Helczynski-et-al_IEEE}L. Helczynski, D. Anderson, R. Fedele, B. Hall, and M. Lisak,
{\it IEEE J. of Sel. Topics in Q. El.}, {\bf 8}, 408 (2002)

\bibitem{Fedele-et-al_PLA1}R. Fedele, P.K. Shukla, M. Onorato, D. Anderson, and M. Lisak,
{\it Phys. Lett. A} {\bf 303}, 61 (2002).

\bibitem{Marklund}M. Marklund, {\it Phys. Plamas} {\bf 12}, 082110
(2005).

\bibitem{Onorato-et-al_PRE}M. Onorato, A. Osborne, R. Fedele, and M. Serio, {\it Phys. Rev. E}
{\bf 67}, 046305 (2003).

\bibitem{Visinescu-Grecu}A. Visinescu and D. Grecu, {\it Eur. Phys. J. B} {\bf 34}, 225 (2003);
A. Visinescu, D. Grecu AIP Conf. Proc. Vol. 729, p. 389 (2004).

\bibitem{Grecu-Visinescu}D. Grecu and A. Visinescu, {\it Rom. J. Phys.} {\bf 50}, nr.1-2 (2005).

\bibitem{Wigner}E. Wigner, {\it Phys. Rev.}, {\bf 40} 749 (1932).

\bibitem{Moyal}J.E. Moyal, {\it Proc. Cambidge Phil. Soc.}, {\bf 45}, 99 (1949).

\bibitem{Ville}J. Ville, {\it Cables et Transmission} {\bf 2}, 61 (1948).

\bibitem{Klimontovich-Silin}Y. Klimontovich and V. Silin,
{\it Sov. Phys. Usp.} {\bf 3}, 84 (1960).

\bibitem{Alber}I.E. Alber, {\it Proc. R. Soc. London}, Ser. A {\bf 636}, 525 (1978).

\bibitem{Landau1}L.D. Landau, {\it Zeitschrift f\"{u}r Physik} {\bf 45}, 430 (1927).

\bibitem{von-Neumann}von Neumann J., {\it Mathematische Grundlagen der
Quantenmechanik} (Springer, Berlin, 1932); {\it Collected Works}
(Oxford, Pergamon, 1963).

\bibitem{Weyl}H. Weyl, {\it Gruppentheorie und Quantenmechanik} (1931); engl.
transl.: {\it The Theory of Groups and Quantum Mechanics} (Dover,
Publ., 1931).

\bibitem{Landau} L.D. Landau, {\it J. Phys. USSR}, {\bf 10}, 25 (1946).

\end{thebibliography}
\end{document}